\documentclass[cits]{myPoS}

\usepackage{macros,amsmath,amstext,amsfonts,amsbsy,amssymb,amscd,epsfig,graphicx}

\PoS{PoS(LAT2005)344}

\title{On the determination of low-energy constants for $\Delta{S}=1$
  transitions\thanks{CERN-PH-TH/2005-175, IFIC/05-45, FTUV-05-1003, BI-TP 2005/41, DESY 05-198}} 

\ShortTitle{On the determination of low-energy constants for
  $\Delta{S}=1$ transitions} 

\author{Leonardo Giusti, Carlos Pena$^\dag$\\
        CERN, Physics Department, Theory Division, CH-1211 Geneva 23,
        Switzerland\\ 
        E-mail: \email{leonardo.giusti@cern.ch},
        \email{carlos.pena.ruano@cern.ch}}

\author{Pilar Hern\'andez\\
        Dpto. F\'{\i}sica Te\'orica and IFIC,\\Edificio Institutos Investigaci\'on, Apt. 22085, E-46071 Valencia, Spain\\
        E-mail: \email{pilar.hernandez@ific.uv.es}}

\author{Mikko Laine\\
        Faculty of Physics, University of Bielefeld, D-33501,
        Bielefeld, Germany\\
        E-mail: \email{laine@physik.uni-bielefeld.de}}

\author{Jan Wennekers, \speaker{Hartmut Wittig}\\
        DESY, Notkestra{\ss}e 85, D-22603 Hamburg, Germany\\
        E-mail: \email{jan.wennekers@desy.de},
        \email{hartmut.wittig@desy.de}}

\abstract{We present our preliminary results for three-point correlation functions
  involving the operators entering the $\Delta{S}=1$ effective Hamiltonian with an active charm quark,
  obtained using overlap fermions in the quenched approximation. This
  is the first computation carried out for valence quark masses small enough
  so as to permit a matching to Quenched Chiral Perturbation Theory in the
  $\epsilon$-regime. The commonly observed large statistical
  fluctuations are tamed by means of low-mode averaging techniques,
  combined with restrictions to individual topological sectors. We
  also discuss the matching of the resulting hadronic matrix elements
  to the effective low-energy constants for $\Delta{S}=1$
  transitions. This involves (a) finite-volume corrections which can
  be evaluated at NLO in Quenched Chiral Perturbation Theory, and (b) the
  short-distance renormalization of the relevant four-quark operators
  in discretizations based on the overlap operator. We discuss
  perturbative estimates for the renormalization factors and possible
  strategies for their non-perturbative evaluation. Our results can be
  used to isolate the long-distance contributions to the $\Delta
  I=1/2$ rule, coming from physics effects around the intrinsic QCD
  scale.}

\FullConference{XXIIIrd International Symposium on Lattice Field
  Theory\\
  25-30 July 2005\\
  Trinity College, Dublin, Ireland}

\begin{document}
\section{Introduction}

A satisfactory understanding of long-standing problems in kaon
physics, such as the well known $\Delta{I}=1/2$ rule, has so far been
elusive. In addition to final-state interactions between the two
pions, the other possible origins of the $\Delta{I}=1/2$ rule are
``intrinsic'' long-distance QCD effects at typical energy scales of a
few hundred \MeV, as well as the decoupling of the charm quark from
the light quark sector, owing to its large mass of around 1.3\,\GeV.
In refs. \cite{strat,charm} we outlined a strategy to identify a
mechanism for the $\Delta{I}=1/2$ rule, by separately quantifying each
of the above contributions. Leaving aside final-state interactions,
our strategy is implemented by computing appropriate hadronic correlation
functions allowing to determine the weak low-energy constants (LECs) appearing in the effective chiral theory. Our
approach is characterized by the following features:
\begin{itemize}
\item The use of overlap fermions \cite{NeubergerDirac} in
  computations of hadronic matrix elements of 4-quark operators
  mediating $\Delta{S}=1$ transitions. As described in \cite{SteLeo01}
  the mixing with operators of lower dimensions usually
  encountered with Wilson fermions is completely avoided.
\item Matching to ChPT in the so-called $\epsilon$-regime of QCD,
  where the chiral counting rules imply that this step can be
  performed at NLO without the appearance of unknown LECs. Since overlap fermions preserve chiral symmetry
  the matching can be performed in a conceptually easy and clean manner at
  non-zero lattice spacing.
\item Investigation of the r\^ole of the charm quark, keeping it as an
  active quark in the formulation of the effective $\Delta{S}=1$
  interaction. This allows to isolate the contributions due to a large
  mass splitting between $m_c$, $m_u$. To this end we start with the
  (unphysical) situation of a mass-degenerate charm quark, $m_c=m_u$
  and compute LECs as a function of
  $m_c$.
\end{itemize}
In this note we demonstrate the feasibility of our strategy in the
mass-degenerate case, $m_c=m_u=m_d=m_s$, where QCD possesses an
$\fourby$ chiral symmetry. Since simulations in the
$\epsilon$-regime are plagued by large statistical fluctuations
\cite{Bietenholz03,lma}, we describe in detail how a reliable signal
can be obtained for 3-point correlation functions using ``low-mode
averaging'' (LMA) \cite{lma,DeG_Sch04}. Furthermore, we discuss the relations
between the computed correlation functions and transition amplitudes
for $K\to\pi\pi$ decays. This requires knowledge of the short-distance
renormalization factors of 4-quark operators, as well as finite-volume
corrections that are computed in ChPT.

\section{$\Delta{S}=1$ transitions with an active charm quark}

In order to make this note self-contained, we report the basic
features of our approach. Ref.\,\cite{strat} can be consulted for full
details. The decay of a neutral kaon into a pair of pions in a state
with isospin~$\alpha$ is described by the transition amplitude
\be
   T(K^0\to\pi\pi\big|_{\alpha}) = i A_\alpha\rme^{i\delta_\alpha},
   \qquad\alpha=0,\,2,
\ee
where $\delta_\alpha$ is the scattering phase shift. The experimental
observation that the amplitude $A_0$ is significantly larger than
$A_2$, i.e. $A_0/A_2 = 22.1$, is called the $\Delta{I}=1/2$ rule. Our
task is the computation of correlation functions involving local
operators, which can be linked to the amplitudes $A_0$ and $A_2$.

The relevant local operators are obtained via the operator product
expansion of the $\Delta{S}=1$ effective weak interaction. For two
generations, the effective weak Hamiltonian with an active charm quark
reads
\bea
& & {\cal{H}}_{\rm w}=\frac{g^2_{\rm{w}}}{4M_W^2}(V_{us})^*V_{ud}
  \sum_{\sigma=\pm}\left\{
       k_1^{\sigma}Q_1^{\sigma}
      +k_2^{\sigma}Q_2^{\sigma} \right\}, \label{eq_Hw_final}\\
& & Q_1^{\pm} = \Big\{
   (\sbar\gamma_{\mu}P_{-}{u})(\ubar\gamma_{\mu}P_{-}{d})
\pm(\sbar\gamma_{\mu}P_{-}{d})(\ubar\gamma_{\mu}P_{-}{u})
   \Big\} - (u\,{\to}\,c),  \label{eq_Q1_bare} \\
& & Q_2^{\pm} = (m_u^2-m_c^2)\Big\{
      m_d({\sbar}P_{+}{d})
     +m_s({\sbar}P_{-}{d}) \Big\},\qquad P_\pm=\half(1\pm\gamma_5).
\eea
Since $Q_2^{\pm}$ does not contribute to the physical
$K\to\pi\pi$ transition we drop it from now on. Note that the
operators $Q_1^{+}$ and $Q_1^{-}$ transform according to
irreducible representations of $\rm SU(4)_L$ of dimensions 84 and 20,
respectively.

The renormalization and mixing patterns of $Q_1^{\pm},
Q_2^{\pm}$ derived formally in the continuum theory are
preserved on the lattice, provided that the lattice Dirac operator~$D$
satisfies the Ginsparg-Wilson relation \cite{GinsWil}, and therefore an exact chiral symmetry at finite lattice spacing exists~\cite{Luscher:1998pq}. If
one furthermore replaces $\psi$ by $\tilde{\psi}=(1-{\half}aD)\psi$,
the resulting local operators in the lattice theory have simple
transformation properties under the chiral symmetry.  Thus, no
mixing with lower-dimensional operators can occur \cite{SteLeo01}.

The amplitudes $A_0$ and $A_2$ can be related to low-energy constants in
an effective low-energy description of $\Delta{S}=1$ weak decays. To
this end we consider the leading order effective chiral Lagrangian
\be
  {\cal L}_{\rm E} = \quarter F^2\,\Tr\left[
  (\partial_\mu U)\partial_\mu U^\dagger \right]
  -\half\Sigma\,\Tr\left[ UM^\dagger\rme^{i\theta/\Nf}
  +MU^\dagger\rme^{-i\theta/\Nf} \right],
\label{eq_Leff_LO}
\ee
where $U\in\rm SU(4)$ denotes the Goldstone bosons, $\theta$ is the
vacuum angle, and $M$ is the quark mass matrix. The LECs $F$ and
$\Sigma$ denote the pion decay constant and the chiral condensate in
the chiral limit. The low-energy counterpart of the $\Delta{S}=1$
effective weak Hamiltonian is obtained at lowest order in the chiral
expansion as
\be
   {\cal H}_{\rm w}^{\rm ChPT} =
   \frac{g^2_{\rm{w}}}{2M_W^2}(V_{us})^*V_{ud}
   \sum_{\sigma=\pm}g_1^\sigma\left\{
   [\widehat{\cal{O}}_1^\sigma]_{suud}-[\widehat{\cal{O}}_1^\sigma]_{sccd}
   \right\},
\label{eq_HwChPT}
\ee
where operators containing $M$ have been neglected, and
\be
  [\widehat{\cal O}_1^{\pm}]_{\alpha\beta\gamma\delta} =
  \quarter F^4
  \left(U\partial_\mu U^\dagger\right)_{\gamma\alpha}
  \left(U\partial_\mu U^\dagger\right)_{\delta\beta} +
  \;\hbox{projections onto irreps. of dim. 84, 20}.
\label{eq_O1_ChPT}
\ee
The expression which links the LECS $g_1^{+}$ and $g_1^{-}$ to the
ratio of amplitudes $A_0/A_2$ at leading order in ChPT then reads
\be
   \frac{A_0}{A_2} = \frac{1}{\sqrt{2}}\left(
   \frac{1}{2}+\frac{3}{2}\frac{g_1^{-}}{g_1^{+}} \right).
\label{eq_A0A2_ChPT}
\ee
Finally, the LECs $g_1^\pm$ can be determined by matching suitable correlation functions in ChPT
and QCD. This leads to
\be
{\frac{g_1^{-}}{g_1^{+}}}\,H =
      \frac{k_1^{-}(M_W/\Lambda)}{k_1^{+}(M_W/\Lambda)}\cdot
      \frac{\widehat{Z}^{-}(g_0)}{\widehat{Z}^{+}(g_0)}\cdot
      \frac{C_1^-}{C_1^+}.
\label{eq_ChPT_QCD}
\ee
Here, the chiral correction factor $H$ is obtained as a ratio of
correlation functions of $[\widehat{\cal O}_1^{\pm}]$ computed in
ChPT, and $C_1^\pm$ are specified in eq.~(\ref{eq:3p}) below. On the RHS the short-distance corrections include the Wilson
coefficients $k_1^\pm$ and the renormalization factors
$\widehat{Z}^\pm$, which relate the unrenormalized operator
$(Q_1^{\pm})_{\rm bare}$, considered at bare coupling $g_0$,
to the renormalization group invariant operator via
\be
    (Q_1^\pm)_{\rm RGI} =
    \widehat{Z}^\pm(g_0)(Q_1^{\pm})_{\rm bare}.
\label{eq_Zhat_def}
\ee
In the following sections we describe the evaluation of the
correlation functions, the chiral correction and renormalization
factors.

\section{Lattice set-up in the SU(4)-symmetric case}

Since preserving chiral symmetry is an essential feature of our setup,
the computation of the correlation functions in
Eqs.~(\ref{eq:2p},\ref{eq:3p}) is performed in an overlap lattice
regularization. In order to match QCD to its effective
low-energy description in the SU(4)-symmetric case, we start by
defining suitable two- and three-point correlation functions of
left-handed currents and four-quark operators in QCD,
namely\footnote{The use of left-handed currents, as explained
in~\cite{Giusti:2002sm,strat}, is particularly convenient for technical reasons.}
\begin{align}
\label{eq:2p}
C(x_0) &= \sum_{\mathbf{x}} \left\langle [J_0(x)]_{\alpha\beta}[J_0(0)]_{\beta\alpha}\right\rangle \, ,\\
\label{eq:3p}
C_1^\pm(x_0,y_0) &= \sum_{\mathbf{x},\mathbf{y}} \left\langle [J_0(x)]_{du}\,Q_1^\pm(0)\,[J_0(y)]_{us}\right\rangle \, ,
\end{align}
where the non-singlet left-handed current $J_\mu$ is defined through
\begin{gather}
[J_\mu(x)]_{\alpha\beta} = (\bar\psi_\alpha\gamma_\mu P_- \tilde \psi_\beta)(x) \, ,
\end{gather}
$\alpha,\beta$ are generic flavour indices, and the replacement
$\psi\to\tilde\psi$ has been performed in the four-quark operators of
Eq.~(\ref{eq_Q1_bare}). Recall that in the SU(4)-symmetric limit the
three-point functions $C_1^\pm$ receive contributions from "figure-8"
diagrams only, since "eye" diagrams exactly cancel due to the
antisymmetrization under $(u \leftrightarrow c)$. It is also useful to
define the following ratios of correlation functions, which will enter
the determination of low-energy constants:
\begin{gather}
\label{eq:r84_r20}
R_{84}(x_0,y_0) = \frac{C_1^+(x_0,y_0)}{C(x_0)C(y_0)} \, ,~~~~~~~~~~~~
R_{20}(x_0,y_0) = \frac{C_1^-(x_0,y_0)}{C(x_0)C(y_0)} \, ,\\
\label{eq:r8420}
R_{84/20}(x_0,y_0) = \frac{C_1^+(x_0,y_0)}{C_1^-(x_0,y_0)} \, .
\end{gather}
Far enough from the location of the source operators, all these ratios
are expected to exhibit plateaux that can be fitted to a constant,
which can then be used as input in the matching procedure to Chiral
Perturbation Theory.

At low quark masses the numerical computation of
correlation functions is usually hampered by the presence of large
statistical fluctuations. The latter can be understood by considering
the expression of the quark propagator in terms of eigenmodes of the
Neuberger-Dirac operator $D$, viz
\begin{gather}
S(x,y) = \frac{1}{V}\sum_k\frac{\eta_k(x)\otimes\eta_k(y)^\dagger}{\bar\lambda_k+m} \, ,
\end{gather}
with $\bar\lambda_k=(1-\half\abar m)\lambda_k$ and
$D\eta_k=\lambda_k\eta_k$.  In the regime $m \lesssim (\Sigma V)^{-1}$, which
allows a matching to Chiral Perturbation Theory in the
$\epsilon$-regime, the low-lying spectrum of $D_m=(1-\half\abar m)D+m$ is discrete with
$\Delta\lambda \approx 1/\Sigma V$, and sizeable contributions to
correlation functions come from a few low modes. Large statistical
fluctuations can be traced back to ``bumpy'' structures in the
wavefunctions of these modes~\cite{Giusti:2003gf,lma}.

In order to treat this problem we use low-mode averaging (LMA)
introduced in~\cite{lma}. The technique proceeds by treating
explicitly the contribution to left-handed quark propagators coming
from a few lowest-lying modes of $D$. To be specific, we split
propagators as
\begin{gather}
\label{eq:prop_split}
S(x,y) = \sum_{k=1}^{n_{\rm low}}\frac{e_k(x)\otimes e_k(y)^\dagger}{\alpha_k} + S^h(x,y) \, ,
\end{gather}
where $S^h$ is the propagator in the orthogonal complement of the subspace spanned by the $n_{\rm low}$ lowest modes,
$e_k = P_\sigma u_k + P_{-\sigma}DP_\sigma u_k$,
$-\sigma$ being the chirality where $D$ possesses zero modes (if any), and $u_k$ is an approximate eigenmode of $D_m^\dagger D_m$:
\begin{gather}
P_\sigma D_m^\dagger D_m P_\sigma u_k = \alpha_k u_k + r_k \, ,~~~~~~~~~~
(u_k,r_l) = 0 ~~~\forall k,l \, .
\end{gather}
After inserting the RHS of Eq.~(\ref{eq:prop_split}) in the
expressions for the correlation functions $C$ and $C_1^\pm$, they can be split as
\begin{gather}
C = C^{ll} + C^{hl} + C^{hh} \, ,\\
C^\pm_1 = C^{\pm;llll}_1 + C^{\pm;lllh}_1 + C^{\pm;llhh}_1 + C^{\pm;lhhh}_1 + C^{\pm;hhhh}_1 \, ,
\end{gather}
where $l$ and $h$ denote the number of ``light'' and ``heavy'' parts
of the quark propagator, respectively. Since the ``light'' part of
$S(x,y)$ is available by construction $\forall x,y$, it is
possible to exploit translational invariance to sample the all-$l$
contributions over many different source points. Furthermore, as
explained in~\cite{lma}, by performing $n_{\rm low}$ additional inversions of
the Dirac operator it is also possible to extend this to the mixed
contribution $C^{hl}$. It is easy to check
that the same applies to the $hlll$ contribution to $C_1^\pm$, as well as to part of the $hhll$ one. As already shown
by the exploratory study in~\cite{Giusti:2004bf}, the application of
this technique with $n_{\rm low}\sim 20$ suffices to obtain a signal
for three-point functions at values of the quark mass of interest in
view of matching to $\epsilon$-regime Chiral Perturbation Theory
results.

\section{Correlation functions in the $\epsilon$ and $p$-regimes}

Our simulation parameters are summarized in Table~\ref{tab:sim_pars}. The simulations for lattice A are those reported in~\cite{strat}, while lattices B and C are new results. The statistics of lattice C is currently being increased. The results quoted for lattices B and C have to be considered preliminary.

\begin{table}[!t]
\begin{tabular}{cccccclr}
\hline
Lattice & $\beta$ & $L/a$ & $T/a$ & $n_{\rm low}$ & $L[{\rm fm}]$ & $~~~~~~~~~~~~~~~~~~~~~~~~~~~am$ & \# cfgs \\
\hline
A & 5.8485 & 12 & 30 & 5 & 1.49 & 0.040,0.053,0.066,0.078,0.092 & 638 \\
B & 5.8485 & 12 & 32 &20 & 1.49 & {\it 0.003,0.005,0.007},0.040 & 681 \\
C & 5.8485 & 16 & 32 &20 & 1.99 & {\it 0.002,0.003},0.020,0.030,0.040,0.060 & $\sim 350$ \\
\hline
\end{tabular}
\caption{Simulation parameters for the runs discussed in the text. The mass values in italics are those corresponding to the $\epsilon$-regime. The statistics indicated for lattice C refers to the number of configurations for masses in the $\epsilon$-regime; for $p$-regime masses the statistics is roughly half of the indicated figure.}
\label{tab:sim_pars}
\end{table}

\begin{figure}[t!]
\vspace{150 truemm}
\begin{center}
\includegraphics{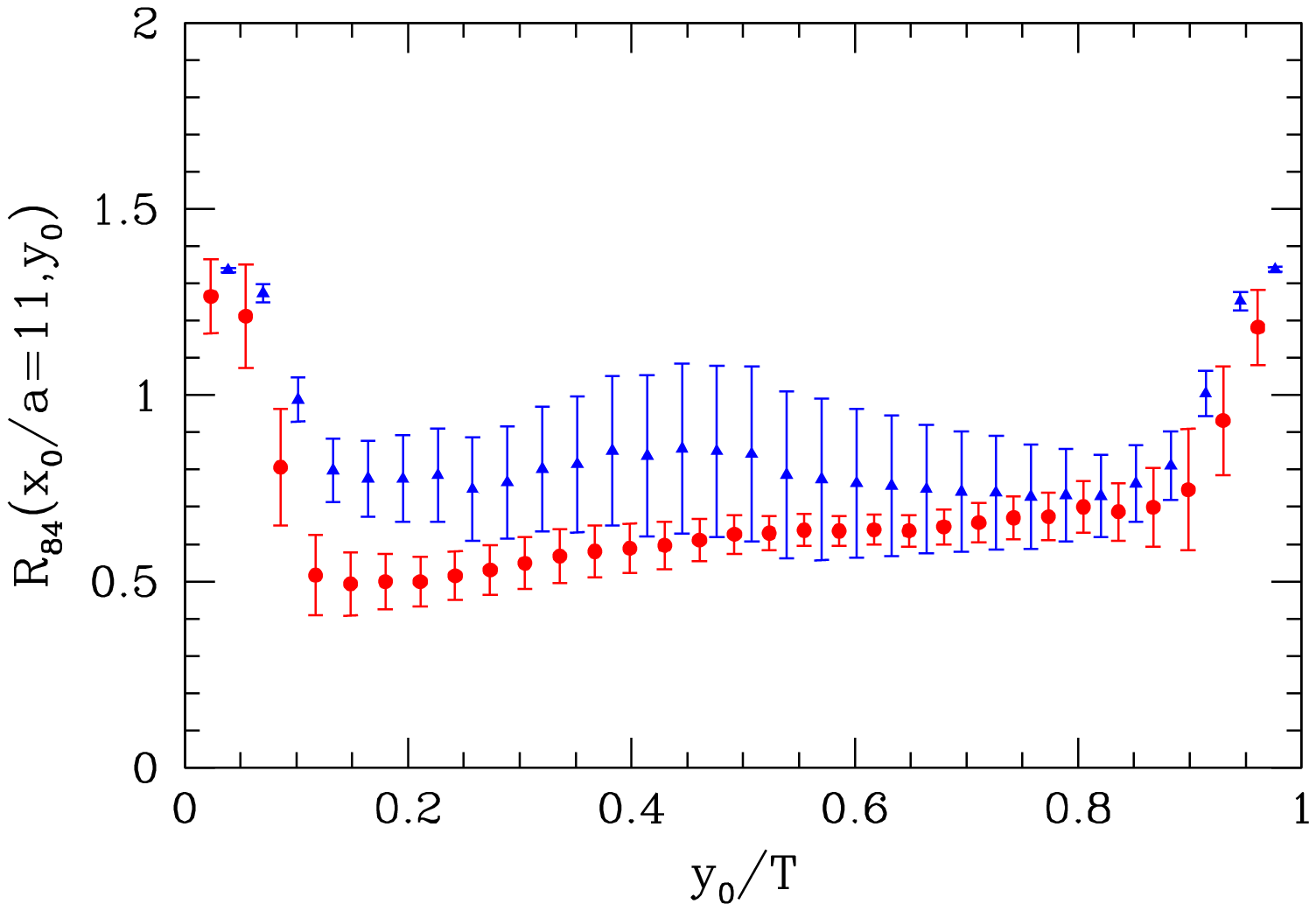}
\includegraphics{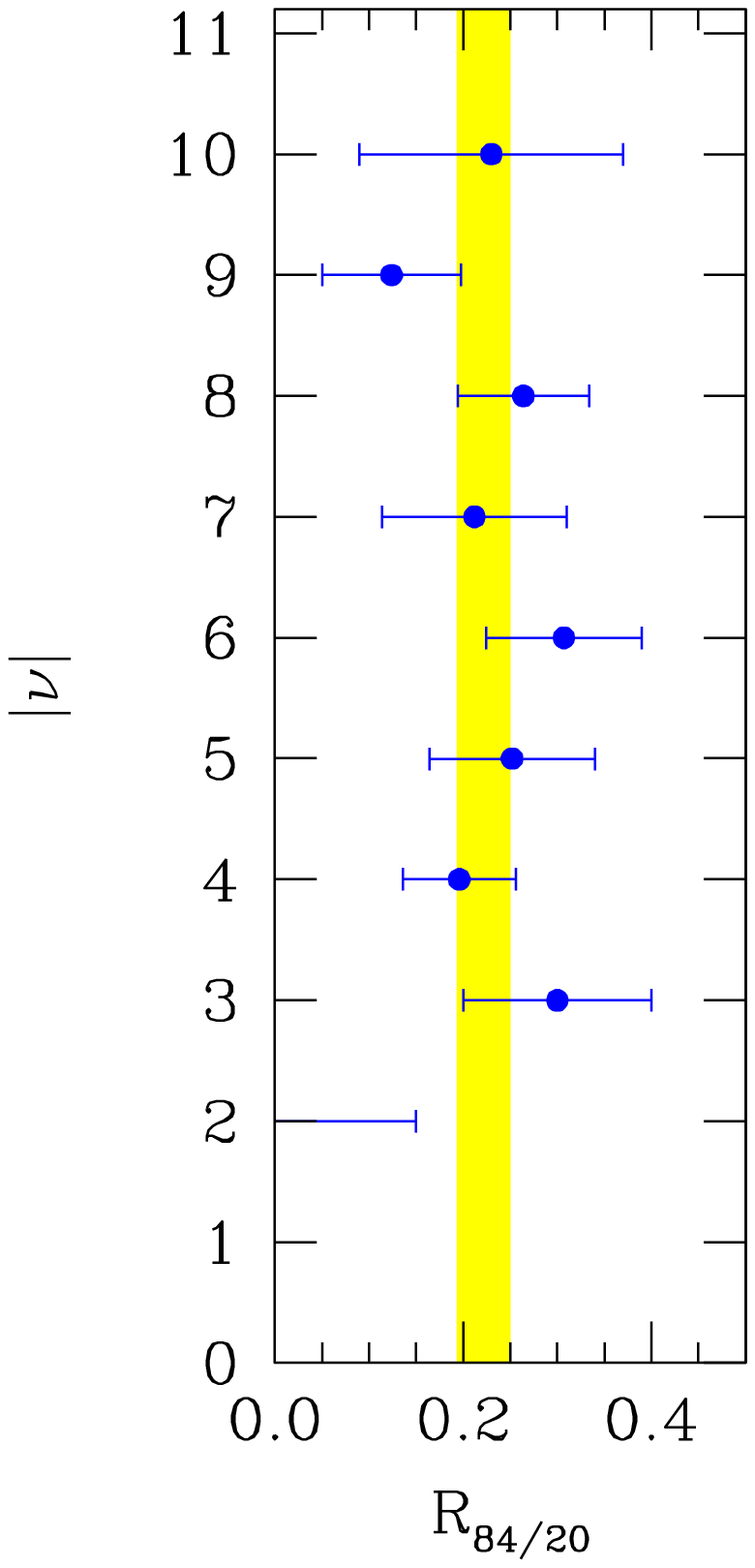}
\end{center}
\vspace{-90 truemm}
\caption{Left: The ratio $R_{84}$ for $am=0.020$, lattice C. The points indicated by circles (triangles) have been obtained with(out) LMA. Right: Weighted average over $|\nu|$ (solid band) of the ratio $R_{84/20}$ for $am=0.003$, lattice C.}
\label{fig:computing_ratios}
\end{figure}

Our main aim is to fit to constants the plateaux in the ratios of correlation functions in Eqs. (\ref{eq:r84_r20},\ref{eq:r8420}). For quark masses in the $p$-regime the procedure is straightforward, and our statistics allows quite precise results for the different ratios. An example for $R_{84/20}$ in the $p$-regime is shown in Fig.~\ref{fig:computing_ratios}. It also shows the effect of LMA on correlation functions at typical $p$-regime masses.

In the $\epsilon$-regime topology plays a special r\^ole~\cite{Leutwyler:1992yt}, and correlation functions are given within fixed topological sectors.
Therefore we proceed by computing the quantities of interest at fixed value $|\nu|$ of the (absolute value of the) topological charge, and then perform a weighted average over $|\nu|$. In order to have large enough statistics within each sector, and taking into account the expected distribution of topological charges, we impose a bound on the largest value of $|\nu|$ entering the average ($|\nu|\le 8$ on lattice B and $|\nu|\le 10$ on lattice C). Furthermore, following the observation that the signal-to-noise ratio in the sectors with lowest $|\nu|$ is poor,\footnote{This can be interpreted as a consequence of the presence of very small eigenvalues of $D$, which in turn induce large statistical fluctuations even after LMA.} for the largest volume (lattice C) we also impose a lower bound $|\nu|\ge 2$. This procedure is illustrated in Fig.~\ref{fig:computing_ratios}. In the $\epsilon$-regime we find no signal at all for the relevant observables if LMA is not implemented. Indeed, to our knowledge, these are the first results for three-point functions obtained at quark masses in the $\epsilon$-regime.

\begin{figure}[t!]
\vspace{150 truemm}
\begin{center}
\includegraphics{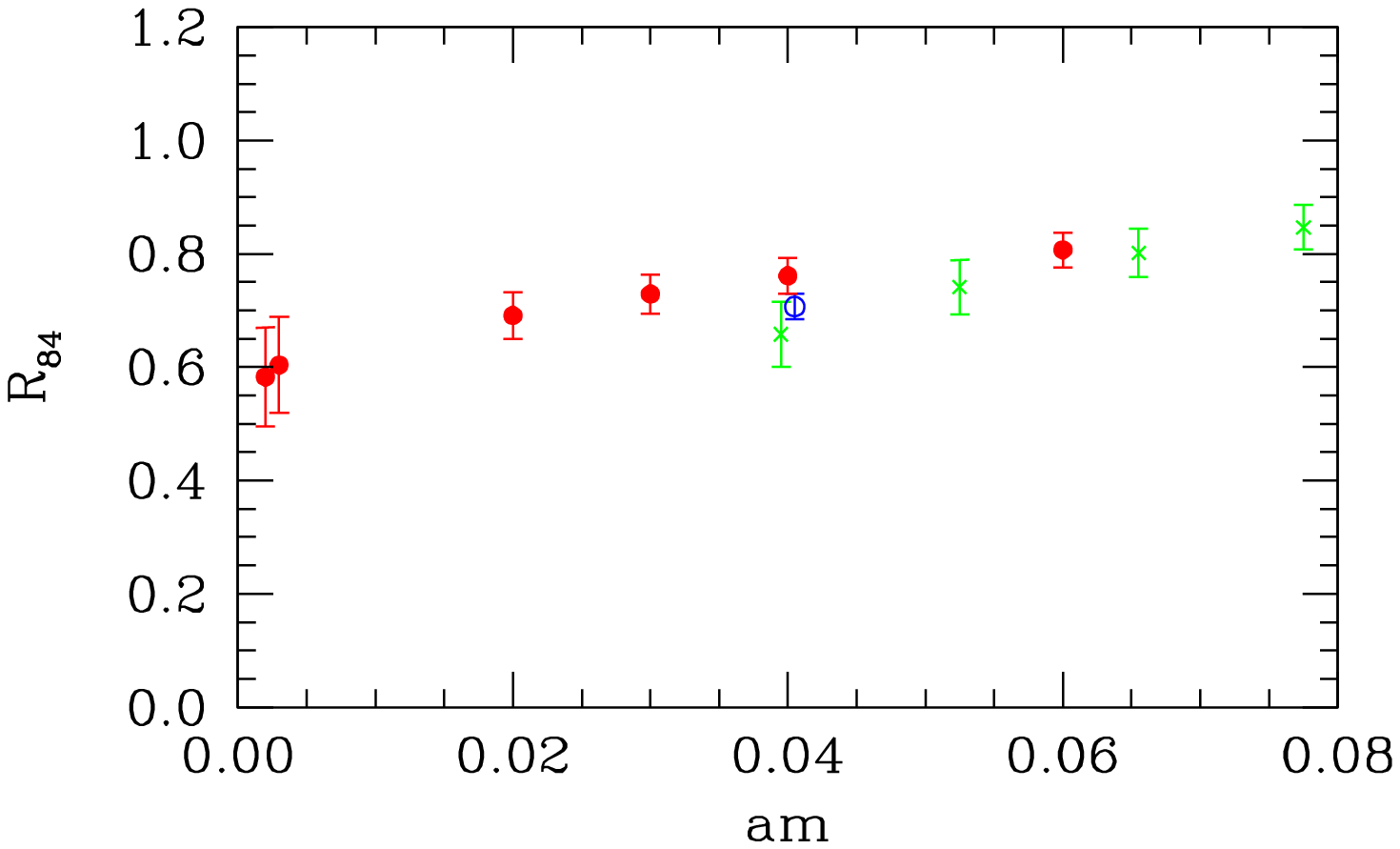}
\includegraphics{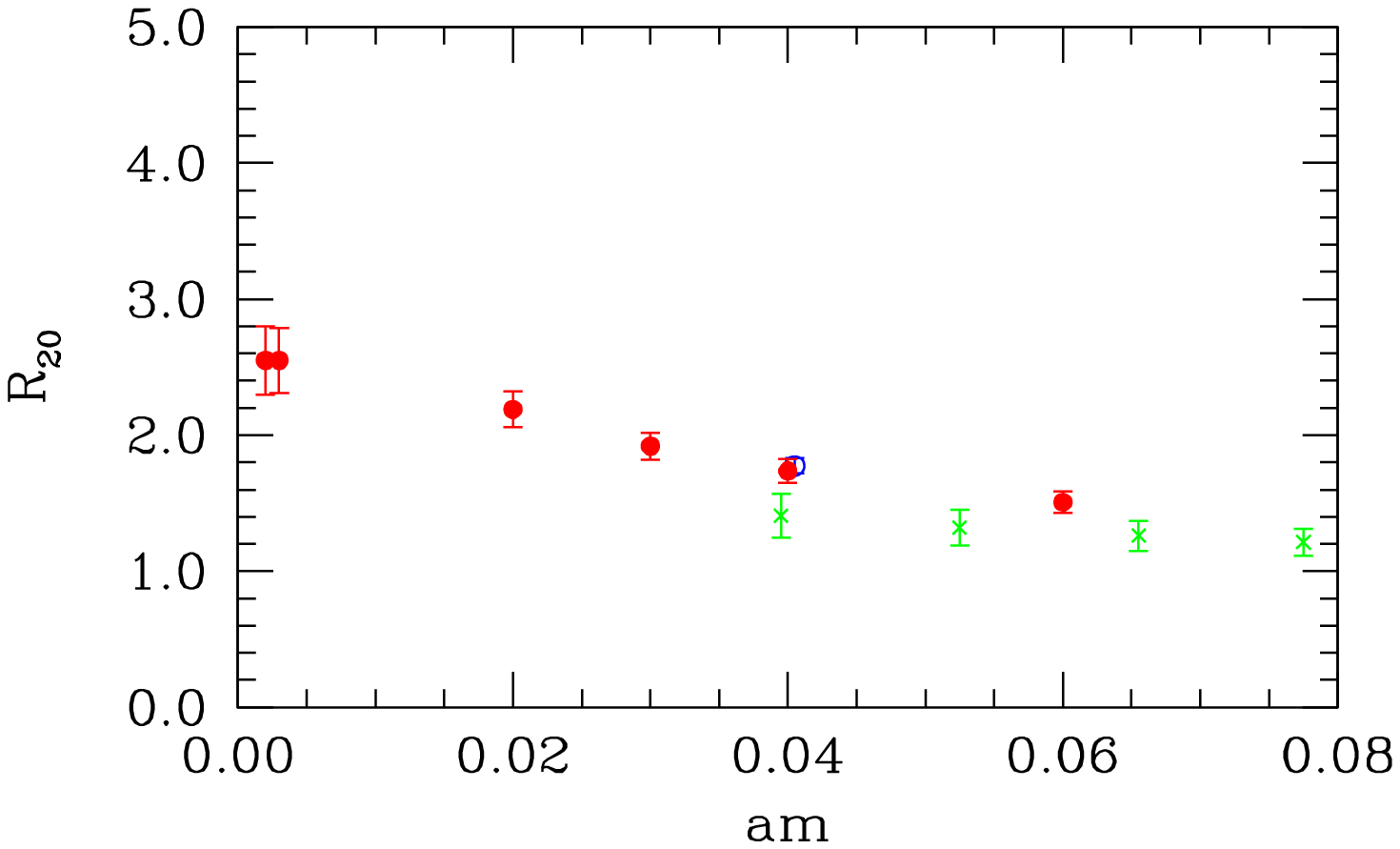}
\includegraphics{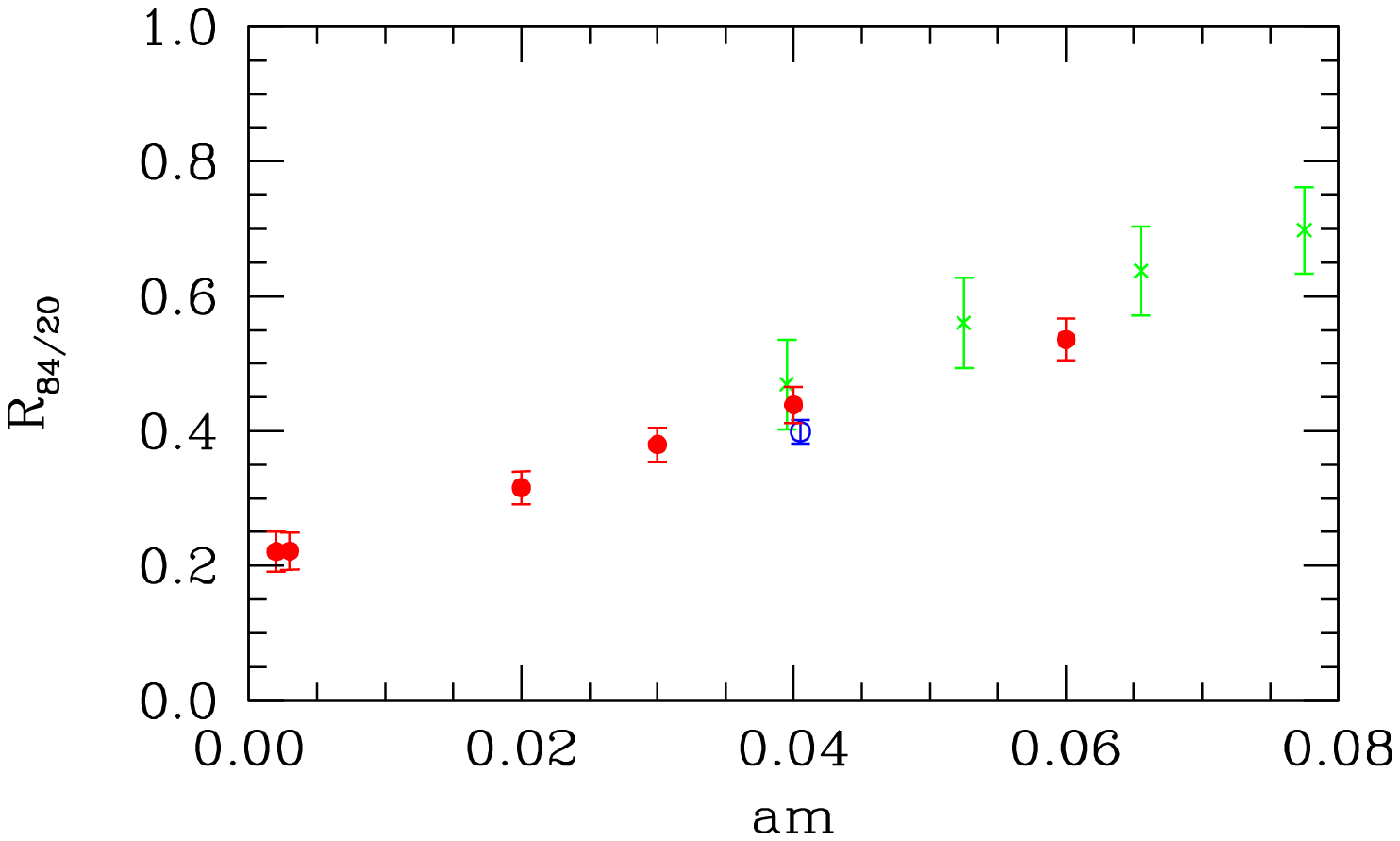}
\end{center}
\vspace{-65 truemm}
\caption{Mass dependence of the ratios $R_{84}$, $R_{20}$ and $R_{84/20}$ for lattices A (crosses), B (empty circles) and C (full circles). Results for lattice C are preliminary. The points at $am=0.040$ have been slightly displaced to improve visibility.}
\label{fig:mass_dep}
\end{figure}

Our most interesting results are those for the mass dependence of the different ratios. They are summarized in Fig.~\ref{fig:mass_dep}, where we put together the $p$-regime results for our three lattices and the $\epsilon$-regime results in our larger volume. Three features worth mentioning are:
Firstly, the mass dependence is remarkably smooth. In particular, there is no strong mass dependence of $R_{84}$ for very small quark masses. Notice, however, that finite volume corrections have to be taken into account for the $\epsilon$-regime points (see below).
Secondly, our results point towards a moderate volume dependence at quark masses corresponding to pseudoscalar meson masses around or below the kaon mass. As far as $R_{84}$ is concerned, this was already observed in~\cite{ALPHA_BK}.
Thirdly, the direct comparison of the different results at $am=0.040$ shows that the effect of LMA with an adequate number of low modes is far from negligible even at moderately large values of the pseudoscalar meson mass.

\section{Renormalization factors for 4-quark operators}

The renormalization factors $\widehat{Z}^\pm(g_0)$ of \eq{eq_Zhat_def}
are scale and scheme independent. For a particular renormalization
scheme $X$ they can be decomposed according to
\be
  \widehat{Z}^\pm(g_0) = c_X^\pm(\mu/\Lambda)\,Z_X^\pm(g_0,a\mu),
\ee
where $\mu$ denotes the renormalization scale, and the coefficients
$c_X^\pm$ are given by
\be
c_{X}^\pm(\mu/\Lambda) =
        (2b_0\gbar^2(\mu))^{\gamma_{\pm;0}/(2b_0)}\,
        \exp\left\{ -\int_0^{\gbar(\mu)}dg
                \left[\frac{\gamma_\pm^{X}(g)}{\beta(g)}
                      +\frac{\gamma_{\pm;0}}{{b_0}g} \right]\right\}.
\ee
The anomalous dimensions $\gamma_\pm^{X}$ are known in perturbation
theory to two loops for several schemes. For
discretizations based on the Neuberger-Dirac operator, the renormalization
factors $\widehat{Z}_{X}(g_0,a\mu)$ have been computed for $X=\rm
RI/MOM$ in perturbation theory at one loop in
ref. \cite{SteLeo01}. Thus, the perturbative renormalization of
suitable ratios of 4-quark operators defined for overlap fermions and
the RI/MOM scheme is
\bea
& & \frac{Z^{-}_{\rm RI}(g_0,a\mu)}{Z^{+}_{\rm RI}(g_0,a\mu)} =
   1+\frac{g_0^2}{16\pi^2}\left\{ 12\ln(4\mu{a})
                                 -2(B_{\rm S}-B_{\rm V}) \right\}
   +\rmO(g_0^4) \label{eq_Zmp_ren} \\
& & \frac{Z^{+}_{\rm RI}(g_0,a\mu)}{Z_{\rm{A;RI}}^2(g_0)} =
   1-\frac{g_0^2}{16\pi^2}\left\{ 4\ln(4\mu{a})
                       -\frac{2}{3}(B_{\rm S}-B_{\rm V}) \right\}
   +\rmO(g_0^4) \label{eq_B84_ren} \\
& & \frac{Z^{-}_{\rm RI}(g_0,a\mu)}{Z_{\rm{A;RI}}^2(g_0)} =
   1+\frac{g_0^2}{16\pi^2}\left\{ 8\ln(4\mu{a})
                       -\frac{4}{3}(B_{\rm S}-B_{\rm V}) \right\}
   +\rmO(g_0^4). \label{eq_B20_ren}
\eea
The coefficients $B_{\rm S}$ and $B_{\rm V}$ are listed in Table~1 of
\cite{SteLeo01}. In addition to 4-quark operators, we have also
considered the renormalization of the axial
current. Eqs. (\ref{eq_B84_ren}) and (\ref{eq_B20_ren}) then serve to
renormalize the corresponding $B$-parameters of the operators
$Q_1^\pm$. The RGI matrix elements are obtained by combining
the above renormalization factors with the coefficients $c_{\rm
RI}^\pm$.

Perturbation theory in the bare coupling $g_0^2$ is known to have bad
convergence properties. The aim of ``mean-field improvement''
\cite{lepenzie93} is to factor out unphysical tadpole contributions in
the perturbative expansion, by a rescaling of the link variable,
$U_\mu(x)\to U_\mu(x)/u_0$. For the Neuberger-Dirac operator defined by
\be
  D_{\rm N}=\frac{\rho}{a}\left(1-A(A^\dagger{A})^{-1/2}\right),
  \qquad \rho=1+s,\quad |s|<1,\quad A=1+s-aD_{\rm w},
\ee
the corresponding rescaling of the quark field is given by
$\psi \to \sqrt{(\rho/\tilde\rho)}\psi,~\tilde\rho
   =(\rho-4)u_0+4$.
For the renormalization factor $Z_{{\cal{O}}_n}=1+g_0^2
z_{{\cal{O}}_n}^{(1)}+\rmO(g_0^4)$ of an $n$-quark operator
${\cal{O}}_n$, the mean-field improved version reads
\be
   Z_{{\cal{O}}_n}^{\rm mfi} =
   \Big(\frac{\rho}{\tilde\rho}\Big)^{n/2} \left\{
   1+\tilde{g}^2\left[
   z_{{\cal{O}}_n}^{(1)}-\frac{n}{2}\frac{\rho-4}{\rho} u_0^{(1)}
   \right]\right\},\quad \tilde{g}^2\equiv
   \frac{g_0^2}{\langle u_0^4\rangle},
\ee
where $u_0=1+u_0^{(1)}g_0^2+\ldots$. When applied to our set of
operators, it is immediately clear that the contributions from the
prefactor $({\rho}/{\tilde\rho})$ as well as those proportional to
$u_0^{(1)}$ drop out in ratios like $Z^{-}/Z^{+}$ and $Z^\pm/Z_{\rm
A}^2$. Mean-field improvement of the expressions in
eqs.~(\ref{eq_Zmp_ren})--(\ref{eq_B20_ren}) is thus simply
accomplished by replacing the bare coupling $g_0^2$ by the
``continuum-like'' coupling $\tilde{g}^2$.

For a reliable determination of operator matrix elements, the use of
non-perturbative estimates for renormalization factors is to be
preferred. The Schr\"odinger functional (SF) offers a general
framework for non-perturbative renormalization of QCD at all scales
\cite{impr:lett}. However, the construction of SF boundary conditions
consistent with the Ginsparg-Wilson relation is quite involved
\cite{taniguchi04}. In ref. \cite{HJLW01} it was therefore proposed to
introduce an intermediate Wilson-type regularization which drops out
in the final result. As an example we now discuss the renormalization
factor $Z^{+}/\za^2$, which is required for the $B$-parameter $B_{\rm
K}$. The desired factor relates the $B$-parameter $B_{\rm K}^{\rm
ov}(g_0)$ computed using overlap fermions, to its RGI counterpart
$\widehat{B}_{\rm K}$. After introducing an intermediate
regularization based, for instance, on twisted mass QCD
\cite{tmQCD}, it can be written as
\be
  \frac{\widehat{Z}^{+}_{\rm ov}}{Z_{\rm A;ov}^2}(g_0) \equiv
    \frac{\widehat{B}_K}{B_K^{\rm ov}(g_0)} =
        \frac{\widehat{B}_K}{ B_K^{\rm tm}(g_0^\prime)}\cdot
        \frac{ B_K^{\rm tm}(g_0^\prime)}{B_K^{\rm ov}(g_0)} 
   =\left[ \lim_{g_0^\prime\to0}
   \frac{\widehat{Z}^{+}_{\rm tm}(g_0^\prime)}{Z_{\rm A;tm}^2(g_0^\prime)}
   \cdot B_K^{\rm tm}(g_0^\prime)\right]
   \cdot 
   \frac{1}{B_K^{\rm ov}(g_0)},
\label{eq_BKren_def}
\ee
where the superscripts ``ov'' and ``tm'' on the unrenormalized
$B$-parameters refer to overlap and twisted mass fermions,
respectively. The key observation is that the expression in square
brackets is nothing but $B_K$ in the continuum limit, which, for
instance, has been computed by the ALPHA Collaboration in quenched QCD
\cite{ALPHA_BK}. Denoting the result by $\widehat{B}_{K}^{\rm ALPHA}$,
the renormalization factor in \eq{eq_BKren_def} is
${\widehat{B}_{K}^{\rm ALPHA}}/{B_{K}^{\rm ov}(g_0)}$. Of course, in
this way one cannot predict $B_K$ any more, since its value is used to
formulate the renormalization condition. However, the procedure can be
used to determine the value of $B_K$ in the chiral limit,
$\widehat{B}_K^\chi$, in units of $\widehat{B}_{K}^{\rm ALPHA}$:
\be
   \widehat{B}_K^\chi = \frac{\widehat{Z}^{+}_{\rm ov}}
                           {Z_{\rm A;ov}^2}(g_0) \times
     B_K^{\chi;\rm ov}(g_0) +\rmO(a^2) =
     \widehat{B}_{K}^{\rm ALPHA}\times
     \frac{B_K^{\chi;\rm ov}(g_0)}{B_{K}^{\rm ov}(g_0)} +\rmO(a^2).
\ee
Note that $B_K^{\chi;\rm ov}(g_0)$ can be obtained from a suitable
ratio of correlators computed in the $\epsilon$-regime, in conjunction
with the appropriate chiral correction factor.

\TABLE[r]{
\begin{tabular}{c c c c}
\hline\hline
\noalign{\vskip0.5ex}
    & bare P.T. & MFI P.T. & non-pert. \\
\noalign{\vskip0.5ex}
\hline
\noalign{\vskip0.3ex}
$\widehat{Z}^{-}/\widehat{Z}^{+}$ & 0.525 & 0.582 & 0.58(8) \\
$\widehat{Z}^{+}/Z_{\rm A}^2$     & 1.242 & 1.193 & 1.20(8) \\
$\widehat{Z}^{-}/Z_{\rm A}^2$     & 0.657 & 0.705 & 0.73(8) \\
\noalign{\vskip0.3ex}
\hline\hline
\end{tabular}
\caption{Perturbative and non-perturbative estimates for RGI
  renormalization factors at $\beta=5.8485$.}
\label{tab_renorm}
}

We now discuss some numerical examples for perturbative and
non-perturbative estimates of renormalization factors. In our
simulations we use $\beta=6/g_0^2=5.8485$. For $\mu=2\,\GeV$ and
$\Lambda=238\,\MeV$ \cite{mbar:pap1}, the perturbative expressions for
the coefficients $c_{\rm RI}^\pm$ yield $c_{\rm RI}^{-}(\mu/\Lambda) =
0.6259$ and $c_{\rm RI}^{+}(\mu/\Lambda) = 1.2735$. Non-perturbative
estimates for the $B$-parameters computed at the physical kaon mass in
the continuum limit of quenched QCD are provided by the ALPHA
collaboration \cite{ALPHA_BK}. The results for ratios of
renormalization factors are listed in Table~\ref{tab_renorm}.

The entries in the table show that non-perturbative estimates for
renormalization factors are remarkably close to perturbative ones. Indeed, even
the differences between perturbative estimates evaluated in ``bare''
or ``mean-field improved'' perturbation theory are small, presumably
since ratios of operators are considered here. This is in stark
contrast to the situation encountered for simple quark bilinears, for
which the deviations between perturbative and non-perturbative
estimates amount to about 30\% at similar values of the bare coupling
\cite{Cond}.

\section{Chiral corrections}

Our strategy of determining the LECs of the $\Delta{S}=1$ weak
interactions requires that the kinematical range where ChPT is
applicable must be accessible to lattice simulations of QCD. The
so-called $\epsilon$-expansion \cite{GasLeut_eps} represents a
systematic low-energy description of QCD in a finite volume for
arbitrarily small quark masses. It is characterized kinematically by
the conditions $m\Sigma{V}\sim\rmO(1)$, $F_\pi{L}\gg1$, where $V=L^4$
is the four-volume, and $m$ is the quark mass. These conditions lead
to different chiral counting rules compared with the more commonly
known $p$-regime. In particular, since the inverse box size counts as
one unit of momentum $L^{-1}\sim\rmO(\epsilon)$, one infers
$m\sim\rmO(\epsilon^4)$ and hence $m_\pi\sim\rmO(\epsilon^2)$. For the
effective Hamiltonian ${\cal H}_{\rm w}^{\rm ChPT}$ of \eq{eq_HwChPT}
this in turn implies that no additional interaction terms are
generated at $\rmO{(\epsilon^2)}$. In other words, the
$\epsilon$-regime allows for a NLO matching of lattice data to ChPT
without the appearance of additional, unknown LECs~\cite{Hernandez:2002ds}.

We can now work out the chiral correction factor $H$ in
\eq{eq_ChPT_QCD}. To this end we define correlation functions of the
left-handed axial current in complete analogy with the fundamental
theory:
\bea
& & {\cal C}^{ab}(x_0) = \int\rmd^3x \left\langle{\cal J}^a_0(x)
   {\cal J}^b_0(0)\right\rangle,\qquad
  {\cal{J}}_\mu^a \equiv \half
  F^2(T^a)_{\alpha\beta}\left(
  U\partial_\mu U^\dagger\right)_{\beta\alpha},
 \label{eq_Cab_def}  \\
& & {[\widehat{\cal C}_1^\pm(x_0,y_0)]}^{ab}_{\alpha\beta\gamma\delta}
   = \int\rmd^3x\int\rmd^3y
   \left\langle
   {\cal J}^a_0(x)[\widehat{\cal
   O}_1^\pm(0)]_{\alpha\beta\gamma\delta}
   {\cal J}^b_0(y).
   \right\rangle, \label{eq_C1pm_ChPT}
\eea
Choosing a diagonal quark mass matrix and flavour matrices $T^a, T^b$
as in eq.\,(D.6) of \cite{strat}, one defines the chiral correction
factor $H$ by
\be
   H\equiv
   \frac{\widehat{\cal{C}}_1^{-}(x_0,y_0)}{\widehat{\cal{C}}_1^{+}(x_0,y_0)}
   = 1-2R(x_0,y_0).
\ee
For later use
we also consider the chiral corrections for $B$-parameters, i.e.
\be
  K^\pm\equiv
  2\,\frac{\widehat{\cal{C}}^{{\sigma}}_1(x_0,y_0)}
        {\widehat{\cal{C}}(x_0)\widehat{\cal{C}}(y_0)}
  =1+{\sigma} R(x_0,y_0),\qquad{\sigma}=\pm.
\ee
Explicit expressions are listed in section\,5.3 of ref.\,\cite{strat}.
In Fig.\,2 of \cite{strat} the quantity $R$ is plotted as a function
of the box size for several lattice geometries. It clearly
demonstrates that chiral corrections are reasonably small for box
sizes $L\geq 1.5\,\fm$ and lattice geometries with $T/L\leq 2$.

\section{Synthesis, conclusions and outlook}

We can now combine our results for ratios of correlation functions
with the appropriate renormalization and NLO chiral correction
factors. We expect that the latter are best controlled for our
dataset ``C'', for which $T/L=2$ and $L=2\,\fm$. The link between the
ratio $g_1^{-}/g_1^{+}$ and the correlation functions is given in
\eq{eq_ChPT_QCD}, while individual values for $g_1^{+}$ are obtained from
\be
   g_1^{+}K^+ = k_1^{+}(M_W/\Lambda)\frac{\widehat{Z}^+(g_0)}{\za^2(g_0)}\cdot
   \frac{4}{3} B_K^{\chi;\rm ov}(g_0),
\ee
and similarly for $g_1^-$. The LECs $g_1^\pm$ are then related to the
amplitudes $A_0,\,A_2$ via LO ChPT.

Our preliminary results for these amplitudes in the SU(4)-symmetric
theory indicate a severe mismatch with experiment: roughly speaking, our value for $A_0$
is too small by a factor~2, while $A_2$ comes out a factor~2 too
large. This produces an estimate for $A_0/A_2$ which is four times
smaller than the one expected from the experimentally observed
$\Delta{I}=1/2$ rule. On the other hand, this is a factor~4 larger than the naive large-$N_{\rm c}$ limit, and does thus  move in the right direction compared with this case.

However, it would be premature to conclude that the $\Delta{I}=1/2$
rule is generated by the decoupling of the charm quark, since the
amplitude $A_2$ is insensitive to the charm mass, yet its experimental
value is not reproduced either in our calculation. Other possibilities
for the observed mismatch are uncontrolled finite-volume corrections,
quenching effects, or even the breakdown of LO ChPT when relating the
LECs to the transition amplitudes. Our future work will thus
concentrate on corroborating our results in the $\epsilon$-regime, as
well as incorporating the effects of a non-degenerate charm quark
mass.  In this context we shall investigate alternative choices of
correlators, which are saturated with zero modes \cite{PP}.

\vskip 5pt
Our calculations were performed on PC clusters at DESY Hamburg, CILEA and the Univer- sity of Valencia, as well as on the
IBM Regatta at FZ J\"ulich. We thank all these institutions and the University of Milano-Bicocca for their support.

\end{document}